\documentclass[]{aa} 
\usepackage{graphicx}
\usepackage{txfonts}
\usepackage{natbib}
\usepackage{lscape}
\usepackage{color}
\bibpunct{(}{)}{;}{a}{}{,}
\newcommand{\sext}{\texttt{SExtractor}}

\newcommand{\re}{$\rm{R_e}$}
\newcommand{\muem}{$\rm{\langle\mu\rangle_e}$}
\newcommand{\delm}{$\rm{\langle\Delta\rangle}$}
\newcommand{\kf}{$\rm{K}$}

\newcommand{\bfilt}{$\rm{B}$}
\newcommand{\vf}{$\rm{V}$}

\newcommand{\ie}{{\em i.e.}}

\begin{document}

   \title{Surface photometry of WINGS\\
   galaxies with GASPHOT}

   \author{M. D'Onofrio\inst{1} \and 
 D. Bindoni\inst{1} \and
 G. Fasano\inst{2} \and
 D. Bettoni\inst{2} \and
 A. Cava\inst{3} \and
 J. Fritz\inst{4} \and 
 M. Gullieuszik\inst{2} \and \\
 P. Kj{\ae}rgaard\inst{5} \and
 A. Moretti\inst{1} \and
 M. Moles\inst{6} \and
 A. Omizzolo\inst{7,2} \and
 B.M. Poggianti\inst{2} \and 
 T. Valentinuzzi\inst{1} \and 
 J. Varela\inst{6}}

   \institute{
     Department of Physics and Astronomy, University of Padova, Vicolo Osservatorio 3, 35122 Padova, Italy \and
     INAF -- Padova Astronomical Observatory, Vicolo Osservatorio 5, 35122 Padova, Italy \and
     Observatoire de Gen{\`e}ve, Universit{\'e} de Gen{\`e}ve, 51 Ch. des Maillettes, 1290 Versoix, Switzerland \and
     Sterrenkundig Observatorium, University of Gent Krijgslaan 281 S9, B-9000 Gent, Belgium \and
     The Niels Bohr Institute for Astronomy Physics and Geophysics, J, Maries Vej 30, 2100 Copenhagen, Denmark \and
     Centro de Estudios de F\'isica del Cosmo de Aragon, Plaza San Juan 1, 44001 Teruel, Spain \and
     Specola Vaticana, 00120 Stato Citt\`a\ del Vaticano
    }

   \date{\today}
 
 \abstract
{}
{We present the B-, V- and K-band surface photometry catalogs obtained
  running the automatic software GASPHOT on galaxies from the WINGS
  cluster survey having isophotal area larger than 200 pixels. The
  catalogs can be downloaded at the Centre de Donn\'ees Astronomiques
  de Strasbourg (CDS).}
{The luminosity growth curves of stars and galaxies in a given catalog
  relative to a given cluster image, are obtained all together, slicing the
  image with a fixed surface brightness step, through several \sext\
  runs. Then, using a single Sersic law convolved with a space-varying
  PSF, GASPHOT performs a simultaneous $\chi^2$ best-fitting of the
  major- and minor-axis luminosity growth curves of galaxies. We outline
  the GASPHOT performances and compare our surface photometry with
  that obtained by \sext, GALFIT and GIM2D. This analysis is aimed at
  providing statistical information about the accuracy generally
  achieved by the softwares for automatic surface photometry of
  galaxies.}
{For each galaxy and for each photometric band the GASPHOT catalogs
  provide the parameters of the Sersic law best-fitting the luminosity
  profiles. They are: the sky coordinates of the galaxy center
  ($R.A.,DEC.$), the total magnitude ($m$), the semi-major axis of the
  effective isophote (\re), the Sersic index ($n$), the axis ratio
  ($b/a$) and a flag parameter ($Q_{FLAG}$) giving a global indication
  of the fit quality. The WINGS-GASPHOT database includes 41,463
  galaxies in the B-band, 42,275 in the V-band, and 71,687 in the
  K-band.  We find that the bright early-type galaxies have larger
  Sersic indices and effective radii, as well as redder colors in their
  center. In general the effective radii increase systematically from
  the K- to the V- and B-band.}
{The GASPHOT photometry turns out to be in fairly good agreement
  with the surface photometry obtained by GALFIT and GIM2D, as well as
  with the aperture photometry provided by \sext. In particular, the
  direct comparison among structural parameters coming from different
  softwares for common galaxies, indicates that the systematic
  differences are in general small. The only significant deviations
  are likely due to the peculiar (and very accurate) image processing
  adopted by WINGS for large galaxies. The main advantages of GASPHOT
  with respect to other tools are: {\it (i)} the automatic finding of
  the local PSF; {\it (ii)} the short CPU time of execution; {\it
    (iii)} the remarkable stability against the choice of the
  initial guess parameters. All these characteristics make GASPHOT an
  ideal tool for blind surface photometry of large galaxy samples in
  wide-field CCD mosaics.}

   \keywords{Surveys - Galaxies : Clusters : General - Catalogs}

   \maketitle

\section{Introduction}
Thanks to the performances of modern CCD detectors and computing
systems, several astronomical surveys have had the chance of mapping
large sky areas, thus providing the opportunity of measuring
photometric and structural properties of thousands of extended sources
using relatively short exposure-time imaging. This has been achieved
also thanks to automatic surface photometry tools offering robust
results and not heavily demanding in terms of computer time. The tools
most widely used for the aperture and surface photometry of galaxies
are \sext\ \citep{BertinArnouts96}, GIM2D \citep{Simard1998}, and
GALFIT \citep{Pengetal2002}.

In the framework of the \textit{Wide-field Imaging of Nearby
  Galaxy-clusters Survey} \footnote{See the WINGS web-site for all the
  details of this project at http://web.oapd.inaf.it/wings.}
\citep[WINGS;][]{Fasan06}, we have devised the tool GASPHOT
\citep[\textit{GAlaxy Surface PHOTometry};][]{PignFas06}, aimed at
performing the automatic surface photometry of large galaxy samples.
The performances of GASPHOT have been already tested on simulated
galaxies and against the results of supervised, single-object
photometry by \citet{PignFas06}.

In the present paper we present the catalogs obtained running GASPHOT
onto the B-, V- and K-band wide-field imaging of the WINGS
survey. Moreover, we compare the results of GASPHOT with those
obtained by \sext, GIM2D and GALFIT. In particular, the comparison
with GIM2D has been done using the catalogs obtained runnig GASPHOT
onto the B-band imaging of the MGC survey \citep{Allenetal2006} and
partly published in \citet{Poggiantietal2013}.

The paper is organized as follows: in Sec.~\ref{Sec1} we recall the
features of the WINGS survey in order to set the context in which the
GASPHOT database is inserted. Section~\ref{Sec2} describes the
guidelines of the software, the data sample analyzed by GASPHOT, and
the typical output files. In Section~\ref{Sec3} the GASPHOT photometry
is compared with that coming from \sext, GIM2D and GALFIT. In
Section~\ref{Sec4}, using the structural parameters coming from
GASPHOT, we present the main scaling relations of the WINGS galaxies
in the different photometric bands. Our conclusions are drawn in
Section~\ref{Sec5}.

\section{The WINGS survey}\label{Sec1}

The WINGS survey \citep{Fasan06} is a long term project especially
designed to provide a robust characterization of the photometric and
spectroscopic properties of galaxies in nearby clusters.  The core of
the survey is WINGS-OPT \citep{Varela09}, that is a set of B- and
V-band images of a complete, X-ray selected sample of 77 clusters with
redshift $0.04<z<0.07$.  The images have been taken with the Wide
Field Camera (WFC, $34'\times 34'$) at the INT-2.5 m telescope in La
Palma (Canary Islands, Spain) and with the Wide Field Imager (WFI,
$34'\times 33'$) at the MPG/ESO-2.2 m telescope in La Silla
(Chile). The optical photometric catalogs have been obtained using
\sext\ and are 90\% complete at V$\sim$21.7, which translates to
$M^{*}_{V} + 6$ at the mean redshift of the survey \citep{Varela09}.
The WINGS-OPT catalogs contain $\sim$400,000 galaxies in both the V-
and B-band. According to \citet[][Table~D.2]{Varela09}, in the
  whole cluster sample the surface brightness limits at
  1.5$\sigma_{bkg}$ ($\sigma_{bkg}$ is the standard deviation per
  pixel of the background) span the ranges 24.7--26.1 (average value:
  25.71) and 25.4--26.9 (average value: 26.39) in the V- and B-band,
  respectively.

\sext\ catalogs have been obtained also for the near-infrared
follow-up of the survey \citep[WINGS-NIR;][]{Valent09}, which consists
of J- and K-band imaging of a subsample of 28 clusters of the
WINGS-OPT sample, taken with the WFCAM camera at the UKIRT
telescope. Each mosaic image covers $\approx$0.79$\rm{deg}^2$. With
the \sext\ analysis the 90\% detection rate limit for galaxies is
reached at J=20.5 and K=19.4. The WINGS-NIR catalogs contain
$\sim$490,000 and $\sim$260,000 galaxies in the K- and J-band,
respectively.  The photometric depth of the WINGS-NIR imaging
  turned out to be slightly worse than that of the WINGS-OPT
  imaging. Thus, for the K-band the surface brightness limit at
  1.5$\sigma_{bkg}$ spans the range 20.6--21.5 \citep[see Table4
  in][average value: 21.15]{Valent09}.

To give a more complete sketch of the observing material available for
WINGS, we just mention that the survey also includes
medium-resolution, multi-fiber spectroscopy (WINGS-SPE) and U-band
photometry (WINGS-UV) of galaxies in subsamples of the WINGS-OPT
cluster sample. We refer to \citet{Cava09} and \citet{omizz} for
details about these follow-ups. Finally, it is worth mentioning that
we are gathering B-, V- and u'-band OmegaCam@VST imaging (one square
degree FOV) of the WINGS clusters in the southern hemisphere (Gullieuszik
et al., in preparation).

Besides the aperture photometry catalogs (\sext) and the surface
photometry catalogs presented here, this huge amount of data has
produced morphological catalogs ($\sim$40,000 galaxies), obtained
using the purposely devised automatic tool MORPHOT \citep{Fasano2012},
and spectroscopic catalogs including redshifts \citep{Cava09}, star
formation histories, stellar masses and ages \citep{Fritz2011}, as
well as equivalent widths and line-indices \citep{Fritz2013} of
$\sim$6,000 galaxies.

A complete description of the WINGS database, including the GASPHOT
catalogs presented here, can be found in \cite{morettietal2014}

\section{The galaxy sample and the GASPHOT catalogs}\label{Sec2}

The surface photometry of galaxies in the WINGS clusters has been
performed on the same sample used for the morphological (MORPHOT)
analysis, \ie\ $\sim$40,000 galaxies with isophotal area larger than
200 pixels at the threshold of 2.5$\sigma_{bkg}$. To handle such a large
number of galaxies the automatic tool GASPHOT was purposely devised.
Details about the software are given in \cite{PignFas06}, together
with tests of the GASPHOT performances, mainly based on simulated
galaxy samples. The code first produces a set of growth curves of
stars and galaxies through several runs of \sext. Then, for each
galaxy, a simultaneous best-fit of the major and minor-axes growth
curves is performed using a single 2D Sersic law convolved with a
space-varying Point Spread Function (PSF). The fitting strategy of
GASPHOT is a sort of hybrid between the 1D equivalent luminosity
profile fitting and the 2D full image fitting technique. Pros and cons
of these approaches are outlined in Sec.\ref{Sec3b6}.

The GASPHOT tool is blind, \ie\ it performs the surface photometry of
all galaxies in a given catalog (relative to a given frame) without
requiring a first guess of the model parameters for each galaxy, as it
occurs for most popular 2D tools. GASPHOT just needs a special care in
the choice of the configuration file parameters that mostly influence
the observed (PSF convolved) light profiles, in particular the
deblending parameter and the detection and analysis threshold
parameters of \sext\ (DEBLEND\_NTHRESH , DETEC\_THRESH and
ANALYSIS\_THRESH), the surface brightness step and the magnitude range
of the stars used to derive the PSF profile \citep{PignFas06}.

After having extracted the major and minor axis growth curves, for each
galaxy the best-fitting procedure provides the total magnitude ($m$), the
axis ratio ($b/a$), the effective radius (\re ), the Sersic index ($n$), and
the $\chi^2$ of the best fit Sersic model. It is worth recalling here that,
since the boundary values used by GASPHOT for $n$ are 0.5 and 8,
finding these output values of the Sersic index is considered as an
indication that the best fitting procedure has been problematic or
unsuccessful.

Although GASPHOT considers the background as a free parameter of
  the best-fitting algorithm, in most cases it is convenient to
  operate with images in which the background has been already roughly
  subtracted (for instance with \sext\ ) and to refine the subtraction
  limiting the range of variability of the background parameter in the
  fitting procedure. \citet{Varela09} describe in detail the procedure
  of background subtraction adopted for the WINGS clusters. Here we
  just recall that the careful modeling and removal of the
  brightest galaxies and stars (most of them equipped with extended
  halos) allows to obtain reliable surface photometry parameters of
  both the bright galaxies themselves and the many small/faint
  companion galaxies usually embedded inside their halos. Such
  procedure obviously results also in a very precise determination of
  the background path to be subtracted from the images. Thus, in our
  case we have allowed GASPHOT to vary the background parameter of
  just 1.8$\times\sigma_{bkg}$.

The CPU time needed to run GASPHOT on a sample of $\sim 600$ galaxies
(the typical number of galaxies in WINGS cluster catalogs) is $\sim$2h
on a server with a double CPU Xeon E5439 @ 2.6GHz (in total 8 cores)
with 16Gb RAM. Most of this time is used by \sext\ to extract the
major and minor axis growth curves.

In the paper of \cite{PignFas06} the output parameters of GASPHOT have
been compared with results of GIM2D and GALFIT using $\sim$15000
simulated galaxies, including multicomponent ($r^{1/4} + exp.$)
galaxies and blended objects, in a wide range of magnitude,
flattening, and radius. It was found that, for objects with threshold
isophotal area greater than 200 pixels, the photometric and structural
parameters derived by GASPHOT are in very good agreement with the
input values of simulations, even for composite luminosity profiles,
blended objects and low surface brightness galaxies. A small number of
outliers were found, but the results were robust in a statistical
sense. The scatter was in general small ($<$15\%), but for single
objects the errors on effective radius \re\ and Sersic index $n$ could
(in a few cases) exceed 20-40\%.  Finally, although giving
similar results on simulated galaxies, GALFIT and GIM2D were found to
be less robust than GASPHOT when using single Sersic model 
  to fit luminosity profiles of real galaxies. The last result relied
on a small sample of galaxies having detailed (visually supervised)
surface photometry.

\begin{figure*}
\centering
\vspace{-3truecm}
\includegraphics[width=16.4cm,clip]{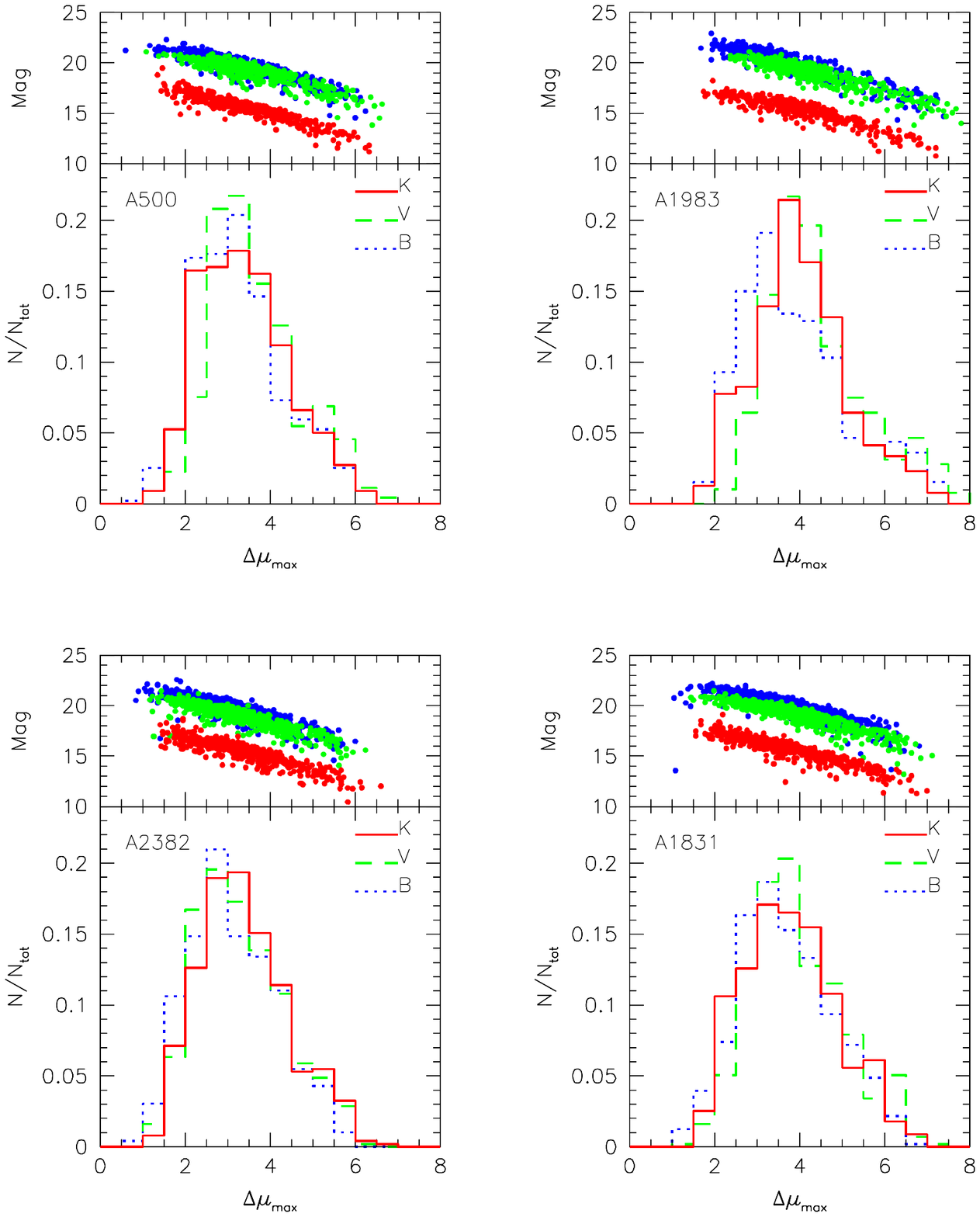}
\vspace{-4truecm}
\caption{Histograms of the photometric depth $\Delta\mu_{max}$
  observed in the \bfilt, \vf\ and \kf\ bands for four different WINGS
  clusters (A550, A1983, A2382 and A1831). The color legend marks the
  various filters. The upper panels above each histogram show the
  total luminosity versus $\Delta\mu_{max}$.}
\label{Fig1}
\end{figure*}

The GASPHOT WINGS-OPT sample in the B(V)-band consists of $41,463$
($42,275$) galaxies of all morphological types detected by \sext\ as
having isophotal area larger than 200 pixels above $2.5\sigma_{bkg}$.
The average number of galaxies per cluster which satisfy the above
condition is $\sim 560$. Using the same criterion, the GASPHOT
WINGS-NIR sample consists of $71,687$ galaxies in the subsample of 28
clusters observed in the K-band \citep{Valent09} ($\sim 2750$ objects
per cluster, on average).  The galaxies in common among the B-, V- and
K-bands are $\sim 10,424$ and belong to 25 clusters\footnote{A119,
  A500, A602, A957x, A1069, A1291, A1631a, A1644, A1795, A1831, A1983,
  A2107, A2124, A2149, A2169, A2382, A2399, A2457, A2572a, A2589,
  IIZW108, MKW3s, RX1022, RX1740, Z8338.}.

For each galaxy, the GASPHOT catalogs list the WINGS identifier and
the best fitting parameters found by GASPHOT for the single Sersic law
model. In particular, the coordinates (RA and DEC) of the center, the
total magnitude, the effective radius, the Sersic index and the axis
ratio (see Table~\ref{tab:Catalogs}). Moreover, for each galaxy,
GASPHOT provides the major- and minor-axis grow curves, as well as the
ellipticity and position angle profiles of the ellipses best fitting
the isophotes. These can be useful to analyze the shape of galaxies,
in particular to test the presence of bars.

Not always the quality of the GASPHOT fit can be judged on the basis of the
$\chi^2$ parameter, since several effects might influence its
value. Among them we mention: the uncertainty on the background value
(in particular close to the very bright objects), the choice of the
deblending parameters of \sext\ for luminosity profile extraction, the
cutting of luminosity profiles for objects close to the CCD borders,
the accuracy of the local PSF and, most of all, the presence of galaxy
substructures that cannot be represented by the Sersic law, especially
when objects are well resolved. On the other hand, the errors provided
by GASPHOT for each output parameter turn out to be usually too small,
since they are just formal uncertainties associated with the fitting
procedure. For these reasons, neither the $\chi^2$, nor the errors on
individual parameters have been included in the catalogs. Instead we
preferred to provide possible users with the global quality index
$Q_{FLAG}$. This is a decimal number corresponding to a binary, 8
digits number.  The first two digits are always set to 0, while the
remaining six are set to 1 when: the Sersic index is equal to 0.5 or 8
(3rd digit), the errors in the estimated parameters (magnitude:4th
digit; effective radius:5th digit; Sersic index:6th digit;
background:7th digit and axial ratio:8th digit) exceed the 98
percentile of the error distributions for the given image. For
instance, $Q_{FLAG}=0$ for good fits, 32 for fits that have Sersic
index equal to 0.5 or 8 (the search interval boundaries used by
GASPHOT for the Sersic index), 2 for fits with too large error on the
background estimation, 16 for fits with too large error on the
estimated magnitude.

The $Q_{FLAG}$ parameter is just an attempt to quantify the problems
encountered during the fit of each galaxy. We believe that, rather
than for single objects, a reliable estimate of the uncertainties of
GASPHOT can be obtained only in a statistical sense. The comparison of
GASPHOT with \sext, GALFIT and GIM2D can provide us with such
statistical uncertainties, thus giving us an idea about the actual
limits of the automatic tools for the surface photometry of galaxies
(see Sec.\ref{Sec3}).

\begin{table}
\centering
\begin{tabular}{lll}
\hline
\hline
Parameter & Units & Description \\
\hline
ID\_WINGS &  NULL & WINGS object identification \\
R.A. &  [deg] & Central right ascension  \\
DEC. &  [deg] & Central declination \\
$m_V$ &  [mag] & Total magnitude \\
\re\ &  [arcsec] & Major axis effective radius \\
$n$ & NULL & Sersic index \\
$b/a$& NULL & Axis ratio \\
$Q_{FLAG}$ & NULL & Quality FLAG \\
\hline 
\end{tabular}
\caption{The parameters provided by GASPHOT for each galaxy in the \vf\ band.} 
\label{tab:Catalogs}
\end{table}

The GASPHOT catalogs refer to fairly homogeneous samples in the
different bands, since they have been obtained from CCD images whose
exposure times have been tuned to reach almost the same photometric
depth. Here, we empirically define the photometric depth as the
interval $\Delta\mu$ between the brightest and the faintest surface
brightness level detected for each galaxy on the CCD image. The above
mentioned homogeneity is illustrated in Figure~\ref{Fig1},
where the $\Delta\mu$ histograms of galaxies in four clusters (two
imaged with INT and two with MPG) for the three bands are plotted as
an example. Of course, larger values of $\Delta\mu$ correspond to
brighter galaxies, but what is noticeable in the figure is that the
range of $\Delta\mu$ is almost the same in the three bands ($\sim
4\div5$mag). Therefore, we are confident that, at least as far as the
photometric depth is concerned, no statistical biases among the
different filters are present in the GASPHOT parameters because of
different galaxy sampling.

In Section~\ref{Sec1} we recall that, with the typical values of
$\sigma_{bkg}$ found in the WINGS-OPT V-band imaging, we obtain an
average value of the isophotal threshold of $\mu_{Thr}(V)\sim 25.7$
mag arcsec$^{-2}$.  It is worth noticing that, at the same signal
to noise level, the images from SDSS in the g- and r-band reach
$\mu_{Thr}(g)<25.2$ mag~arcsec$^{-2}$ and $\mu_{Thr}(r)<24.7$
mag~arcsec$^{-2}$, respectively.

\section{Internal and external comparisons}\label{Sec3}

\citet{PignFas06} checked the GASPHOT performances against the results
coming from detailed, single-object surface photometry of 231 early
type galaxies published by \cite{Fasano2003} and against the surface
photometry parameters obtained by \cite{Smail1997} from HST imaging of
galaxies in the cluster Abell~370. They found a generally good
agreement between automatic (GASPHOT) and single-object surface
photometry parameters, although a large scatter and a slight tendency
of underestimating the total luminosity and the effective radius of very
large galaxies seemed to be present in GASPHOT with respect to the
supervised, single-object surface photometry.

In this section we test the GASPHOT results against \sext\ and GALFIT,
using the WINGS data, and against GIM2D, using the Padova Millenium
Galaxy and Group Catalog PM2GC \citep{Calvi}. In each comparison we
use only galaxies with \sext\ flag equal to zero and found by \sext\
as having threshold area (above 2.5$\sigma_{bkg}$) greater than 200
pixels. Moreover, we decided to exclude from this comparison those
galaxies for which GASPHOT gives $n$=0.5 or $n$=8 (boundary values for
the search $n$ interval), as well as those for which, according to
GASPHOT, the average surface brightness within the effective isophote
turns out to exceed 21.5, 25.5 and 26.5 for the K-, V- and B-band,
respectively. In fact, according to the quality index $Q_{FLAG}$,
beyond these values the surface photometry parameters provided by
GASPHOT for our galaxy samples becomes largely unreliable.

\subsection{GASPHOT vs. \sext}\label{Sec3a}

\begin{figure*}
\resizebox{\hsize}{!}{\includegraphics{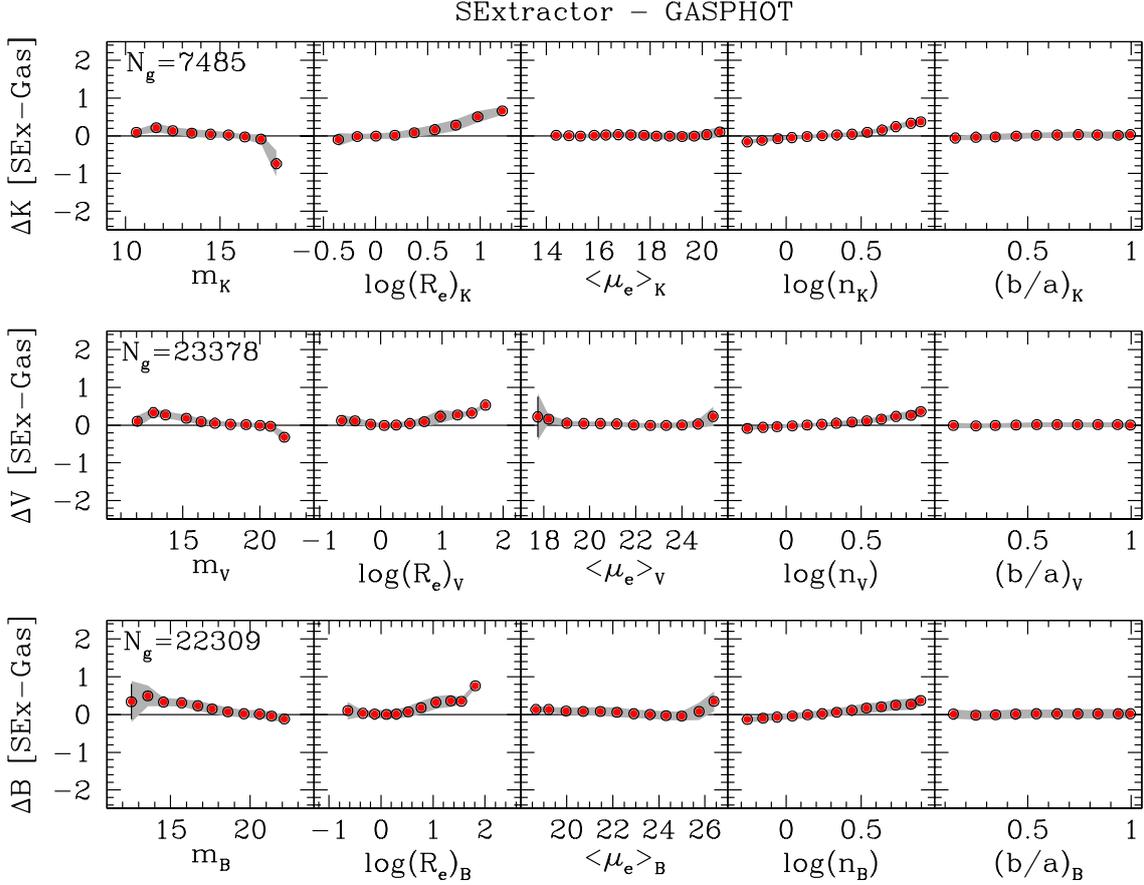}}
\vspace{-1.5truecm}
\caption{Median total magnitude differences between \sext\ and GASPHOT
  in the K-, V- and B-band (top to bottom) binned as a function of the
  absolute magnitude, the effective radius and surface brightness, the
  Sersic index and the axis ratio derived by GASPHOT (left to
  right). The error bars represent the uncertainties of the median
  values in each bin, while the shaded bands give the semi-inter
    quartile ranges of the distributions of the deviations}. In this
  case, both quantities are very small (comparable with the size of
  the points). The sizes of the samples used for the comparisons are
  reported in the leftmost panel for each filter
\label{Fig2}
\end{figure*}

In Figure~\ref{Fig2} the median total magnitude differences
$\Delta$m between \sext\ and GASPHOT in the K-, V- and B-band (top to
bottom) are binned as a function of the best fit quantities derived by
GASPHOT.  They are (left to right): the absolute magnitude, the
effective radius and surface brightness, the Sersic index and the axis
ratio. The error bars represent the $r.m.s$ uncertainty of the median
values in each bin. The number of galaxies used for the comparisons
are 7485, 23378 and 22309 for the K-, V- and B-band, respectively.

The most evident feature in Figure~\ref{Fig2} is the dependence of
$\Delta$m on galaxy size in all wavebands. For large galaxies, the
\sext\ magnitudes turn out to be fainter than the GASPHOT ones. We have
verified such a trend to be particularly evident for late-type
galaxies in the B-band. This depends on the fact that, in spite of the
accuracy in choosing the deblending parameters, in many cases \sext\
tends to erroneously split large spirals into multiple, smaller
objects (HII regions and other light blobs). Instead, in the case of
early-type galaxies, the magnitude difference, still present and
positive, reflects the well known inability of \sext\ to extrapolate
the smoothly decreasing (high Sersic index) outer profiles of galaxies
\citep{France98}. This is confirmed by
the smooth rise of $\Delta$m as a function of the Sersic index in all
wavebands (see Figure~\ref{Fig2}), as well as by the attenuation of
the bias when the first three ranked most luminous galaxies in each
cluster are removed from the WINGS sample. The dependence of $\Delta$m
on galaxy size also determines the behavior of $\Delta$m as a
function of the luminosity. Instead, no dependence at all of $\Delta$m
on the axial ratio is found.

\subsection{GASPHOT vs. GALFIT and GIM2D}\label{Sec3b}

\begin{figure*}
\resizebox{\hsize}{!}{\includegraphics{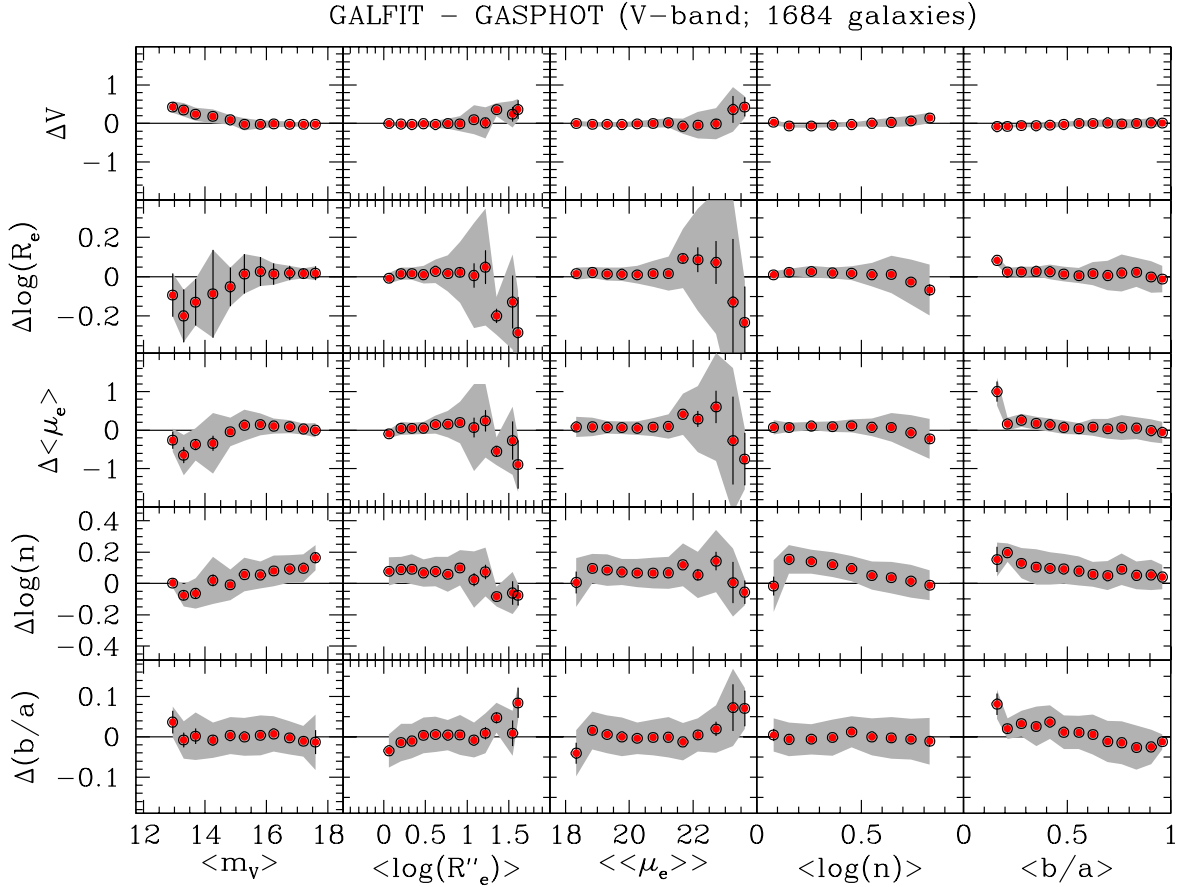}}
\vspace{-1.5truecm}
\caption{Comparison between the results of GASPHOT and GALFIT surface
  photometry for 1684 galaxies randomly extracted from the WINGS
  V-band catalogs. The comparison is made for apparent total
  magnitude, effective radius in arcseconds, effective average surface
  brightness, Sersic index and axis ratio. The above parameters in
  abscissa are averaged between the tools under comparison, while the
  differences between the values found by the tools are reported for
  each parameter on the ordinate, binned over the whole set of
  parameters. As in Fig.~\ref{Fig2}, the error bars represent the
  uncertainties of the median values of the differences in each bin,
  while the shaded bands give the semi-inter quartile ranges of
    the distributions of the deviations.}
\label{Fig3}
\end{figure*}

\begin{figure*}
\resizebox{\hsize}{!}{\includegraphics{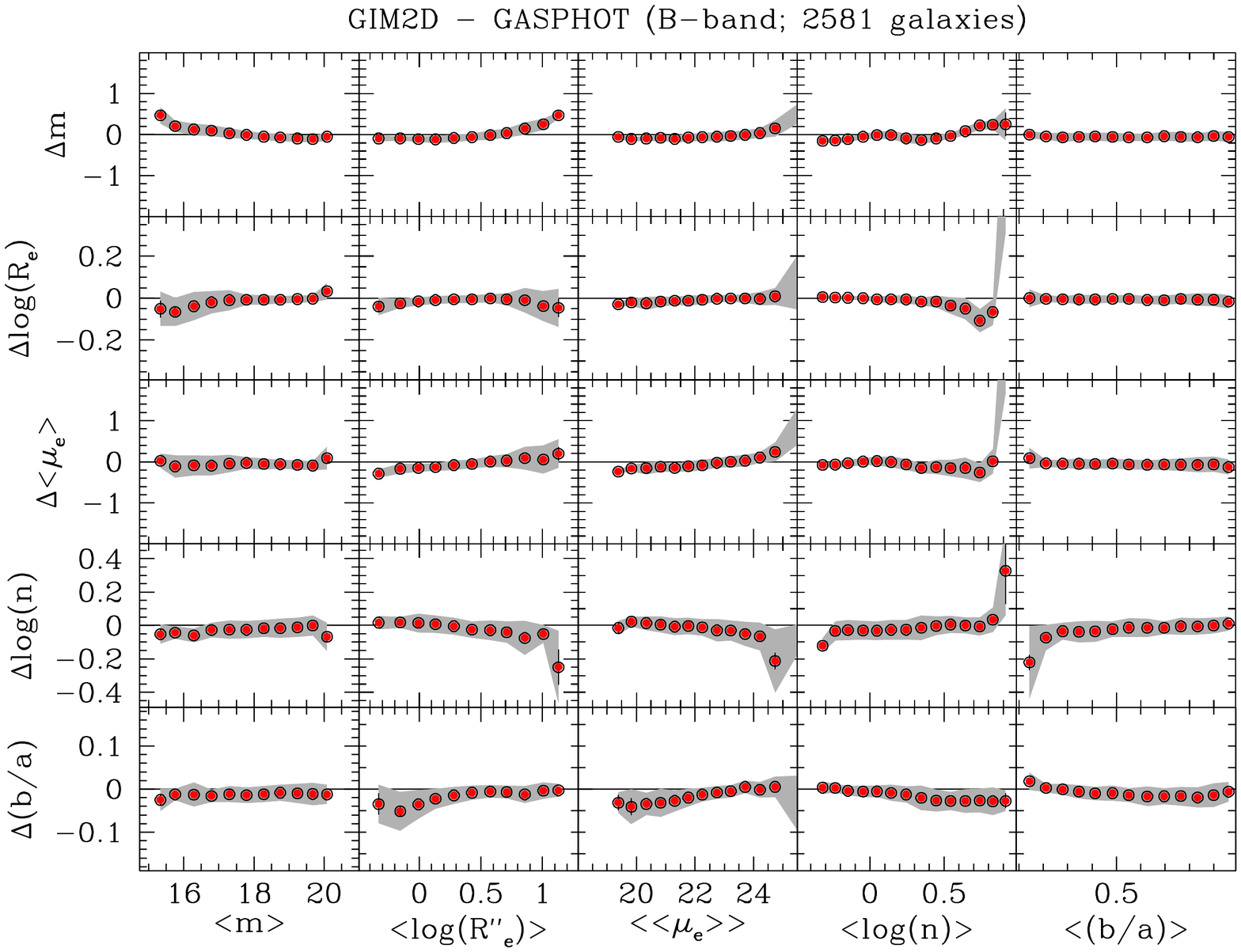}}
\vspace{-1.5truecm}
\caption{Same as Fig.~\ref{Fig3}, but for the GIM2D - GASPHOT
  comparison (B-band) on the sample of 2581 galaxies in common between
  the surveys PM2GC and MGC \citep{Allenetal2006}.}
\label{Fig4}
\end{figure*}

\citet{PignFas06} showed that the performances of GASPHOT
on artificial galaxies are similar to those of GALFIT and GIM2D for
large and regular galaxies, while for automatic surface photometry of
small galaxies and blended objects, GASPHOT provides more robust
results than GALFIT and GIM2D (see Cols. 6 and 9 of Table 1 and
Figs. 12 and 13 in \citet{PignFas06}). This is a crucial feature when
dealing with blind surface photometry of huge galaxy samples.

In this section, we use real galaxies to perform the comparison
between the automatic surface photometry parameters from GASPHOT and
from the above mentioned tools GALFIT and GIM2D. For both comparisons
(GALFIT-GASPHOT and GASPHOT-GIM2D) the model used to fit the galaxy
luminosity profiles was the single 2D Sersic law, with constant
ellipticity and position angle.

\subsubsection{Samples used for the comparisons}\label{Sec3b1}

For the comparison between GASPHOT and GALFIT, we use a sample of 1684
galaxies randomly extracted from the WINGS-GASPHOT catalogs among
those having $Q_{FLAG}=0$ and belonging to clusters whose V-band
WINGS-OPT imaging has been obtained in good seeing conditions, 
  with minimal PSF variation over the cluster field. On this galaxy
sample GALFIT has been run taking as initial guess for the parameters
the V-band photometric and geometric quantities provided by the
WINGS-OPT \sext\ catalogs.

The CPU time required by GALFIT to produce the surface photometry
  parameters of a single galaxy turned out to be (on average) about 2
  times longer than in the case of GASPHOT. This is likely because
  GALFIT has to handle the whole set of pixels belonging to each
  galaxy, while GASPHOT just deals with the major and minor axis
  growth curves.

  The comparison between GIM2D and GASPHOT has been performed using
  galaxies in a sub-sample of the Millennium Galaxy Catalog
  \citep[hereafter MGC;][]{Liske,Cross}. The MGC survey is based on \bfilt\
  band imaging taken with the WFC camera of the Isaac Newton Telescope
  (the same used by WINGS in the northern hemisphere; pixel size of
  0.333 arcsec) along an equatorial strip covering an area of
  $\sim37.5$ deg$^2$. The MGC images reach an isophotal detection
  limit of $26.0$ mag arcsec$^{-2}$.

  The GIM2D data come from three different works based on the MGC
  imaging: the surface photometry by \cite{Allenetal2006}, that of the
  New York University Value Added Catalogue
  \citep[NYUVAC][]{Blantonetal2005}, and that from the SLOAN DR7 data
  \citep{Simardetal2011}.

  We have obtained GASPHOT surface photometry for a sample of galaxies
  in the PM2GC \citep{Calvi}, that is a galaxy catalog extracted from
  the MGC and representative of the general field population in the
  local Universe ($0.04 \leq z \leq 0.1$).

  A preliminary comparison between GASPHOT and GIM2D was presented by
  \cite{Poggiantietal2013} for 618 galaxies in common between the
  PM2GC and MGC surveys. \cite{Poggiantietal2013} found that the
  agreement between GASPHOT and GIM2D is generally good, with a
  tendency for the GASPHOT radii to be slightly larger than the
  others. The median difference between the effective radii \re\ is
  about $0.03\pm0.04$ dex with respect to the data of
  \cite{Allenetal2006}, $0.03\pm0.06$ dex with respect to the NYUVAC,
  and $-0.01\pm0.04$ dex with respect to \cite{Simardetal2011}.

  In the present paper, the comparison between GASPHOT and GIM2D has
  been done using an extended sample of 2581 galaxies in common
  between the PM2GC and MGC surveys.

  Since for this comparison we use GIM2D literature data, in this case
  we are not allowed to directly compare the average CPU time required
  by the tools to produce the surface photometry parameters of a
  single galaxy. However, it is worth recalling that GIM2D was found
  by \citet{PignFas06} to be significantly more expensive in terms of
  CPU time with respect to both GASPHOT and GALFIT.

To discuss in detail the features of GASPHOT, GALFIT and GIM2D is
  beyond the scope of this paper. The reader can found them in the
  above cited papers. Neither we aim here to propose a ranking of
  goodness for the three tools. The comparison we are going to perform
  in this section is just intended at providing an estimate of the
  uncertainties of the surface photometry parameters obtained with
  automatic tools.  Still, it can be useful to summarize here how the
  data used for the comparison have been obtained, in particular in
  the matter of the sky background subtraction, the PSF modeling and
  the best-fitting scheme.

\subsubsection{Sky subtraction}\label{Sec3b2}

The WINGS images used here for both GASPHOT and GALFIT surface
  photometry have already been sky subtracted using the procedure
  described in \citet[][see also Section 2]{Varela09}.

  GASPHOT can use the sky level $I_{bkg}$ as a free model-parameter of
  the best-fitting procedure. However, since this could be dangerous,
  particularly for blended objects, the user can limit the range of
  variation for this parameter when one is confident that a careful
  sky subtraction has already been performed on the images. This is
  our case and we allowed $I_{bkg}$ to vary of
  1.8$\times\sigma_{bkg}$ at most.

  GALFIT can also consider the sky level $I_{bkg}$ as a free
  model-parameter of the best-fitting procedure. However, since no
  restricted range of $I_{bkg}$ variation is allowed in GALFIT, we
  preferred not to include it among the free parameters, fixing its
  value at zero.

  The treatment of the background subtraction is not homogeneous in
  the literature sources of the GIM2D surface photometry data used for
  the present comparison
  \citep{Blantonetal2005,Allenetal2006,Simardetal2011}. It ranges from
  a crude \sext\ estimate of the global sky level to a more accurate
  determination, adopting for each galaxy a minimum distance of
  background pixels from object pixels, defined by segmentation mask
  images.

  \subsubsection{PSF modeling}\label{Sec3b3}

GASPHOT automatically extracts the profiles of the stars, models the
  variation of the FWHM through the field with a 2D polynomial of
  user-defined degree, and combines the PSF profiles, after having
  re-scaled them according to the space varying model obtained
  previously. Finally, a multi-Gaussian function is used to perform
  the $\chi^2$ best-fitting of the average PSF profile. The Sersic
  profiles are then convolved with a PSF whose gaussian coefficients
  depend on the position of the galaxy in the frame.

  Both GALFIT and GIM2D assume the user to be able to provide for each
  galaxy a suitable PSF, both from star images or by functional form.
  When running GALFIT on our sample of WINGS galaxies, we decided to
  adopt a single average PSF image for each cluster, so we could not
  account for minor PSF differences over the image. However, due to
  the previously outlined choice of the WINGS imaging for the
  GASPHOT-GALFIT comparison, this should not significantly contribute
  to worsen the results.

  Again, in the case of GIM2D the determination of the PSF is not
  homogeneous in the literature data used for the present comparison,
  ranging from a unique PSF for all galaxies in a given image to a
  more sophisticated (space varying) treatment, as well as from a
  simple gaussian profile to a more complex (functional or
  user-defined) form.

\subsubsection{Best-fitting}\label{Sec3b4}

GALFIT uses the Levenberg-Marquardt downhill-gradient method to
  derive the best fit.  An error map image is automatically produced
  by the software. At each pixel position the Poisson error is evaluated
  on the basis of the gain and read-noise parameters contained in the image header. Good
  fits can be obtained only when the error map is well known and
  used as weighting image.

  The Metropolis best fitting algorithm used by GIM2D
  \citep{metropolis53,saha94} is more CPU expensive than the
  Levenberg-Marquardt downhill-gradient method used by
  GALFIT. However, it is claimed to be particularly suited to explore
  a n-dimensional parameter space (with n possibly larger than 10)
  having a very complicated topology with local minima at low $S/N$
  ratios. As in GALFIT, a noise map is used to weight pixels.

  After having produced the isophotes of all galaxies together (see
  Section~{Sec2}), GASPHOT performs, for each galaxy, a simultaneous
  Levenberg-Marquardt $\chi^2$ best-fit of the major and minor axis
  growth curves with a 2D Sersic law, convolved with the proper
  PSF. Each point of the growth curves is weighted according to the
  statistical uncertainties on both the integrated isophotal magnitude
  and the radius (pixellation). With respect to the $S/N$ driven,
  pixel-by-pixel weighting, commonly used in the genuine 2D tools,
  this procedure tends to overweight the outer part of the profiles,
  being less sensitive to the high-$S/N$ peculiar features
  (dust-lanes, cores, small bars and rings, etc...) often affecting
  the innermost galaxy body. This is likely to make GASPHOT
  particularly useful when dealing with large/huge galaxy samples, for which
  detailed, single-object (visually supervised) modeling must be
  sacrificed to the advantage of speed and robustness.

\subsubsection{Results}\label{Sec3b5}

\begin{table}
\centering
\begin{tabular}{cc|c|c|c|c|c}
\hline
\multicolumn{7}{c}{GALFIT-GASPHOT} \\
\hline
\hline
 & & $\Delta$m & $\Delta$log\re & $\Delta$\muem & $\Delta$log$n$ & $\Delta$(b/a) \\
\hline
&&&&&&\\ 
& \delm & -0.020 & 0.016 & 0.086 & 0.077 & -0.001 \\
&& & & & & \\
& $\sigma_\Delta$ &  0.123 & 0.077 & 0.320 & 0.116 & 0.067 \\
&&&&&&\\ 
\hline
\multicolumn{7}{c}{GIM2D-GASPHOT} \\
\hline
&&&&&&\\ 
& \delm & -0.068 & -0.006 & -0.049 & -0.014 & -0.012 \\
&& & & & & \\
& $\sigma_\Delta$ &  0.125 & 0.030 & 0.187 & 0.080 & 0.028 \\
&&&&&&\\ 
\hline 
\end{tabular}
\caption{Global median values and $r.m.s.$ of the
  differences (GALFIT-GASPHOT) and (GIM2D-GASPHOT) are reported for the
  apparent magnitude (V- and B-band, for the first and second
  comparison, respectively), the effective radius and surface
  brightness, the Sersic index and the axis ratio.} 
\label{tab:Differences}
\end{table}
 
\begin{figure}[h]
\resizebox{\hsize}{!}{\includegraphics{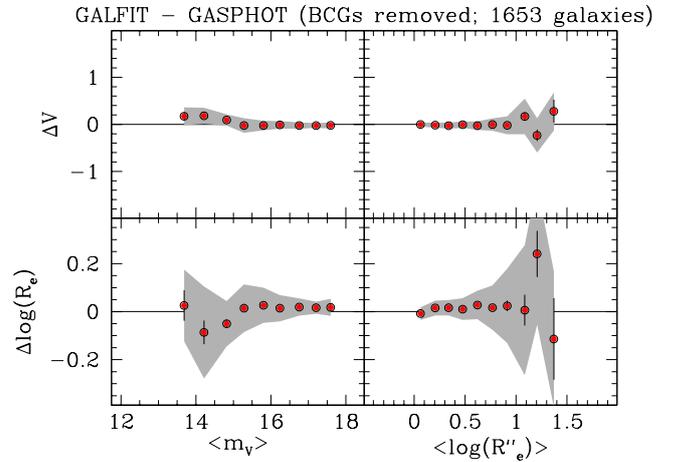}}
\caption{Apparent magnitude and effective radius comparison between
  GASPHOT and GALFIT for the sample of Fig.~\ref{Fig3}, after
  removal of the BCGs. In this case the size-driven bias visible in
  Fig. ~\ref{Fig3} turns out to be much lower or even absent (see
  text).}
\label{Fig5}
\end{figure}

\begin{figure*}[h]
\vspace{-2truecm}
\resizebox{\hsize}{!}{\includegraphics{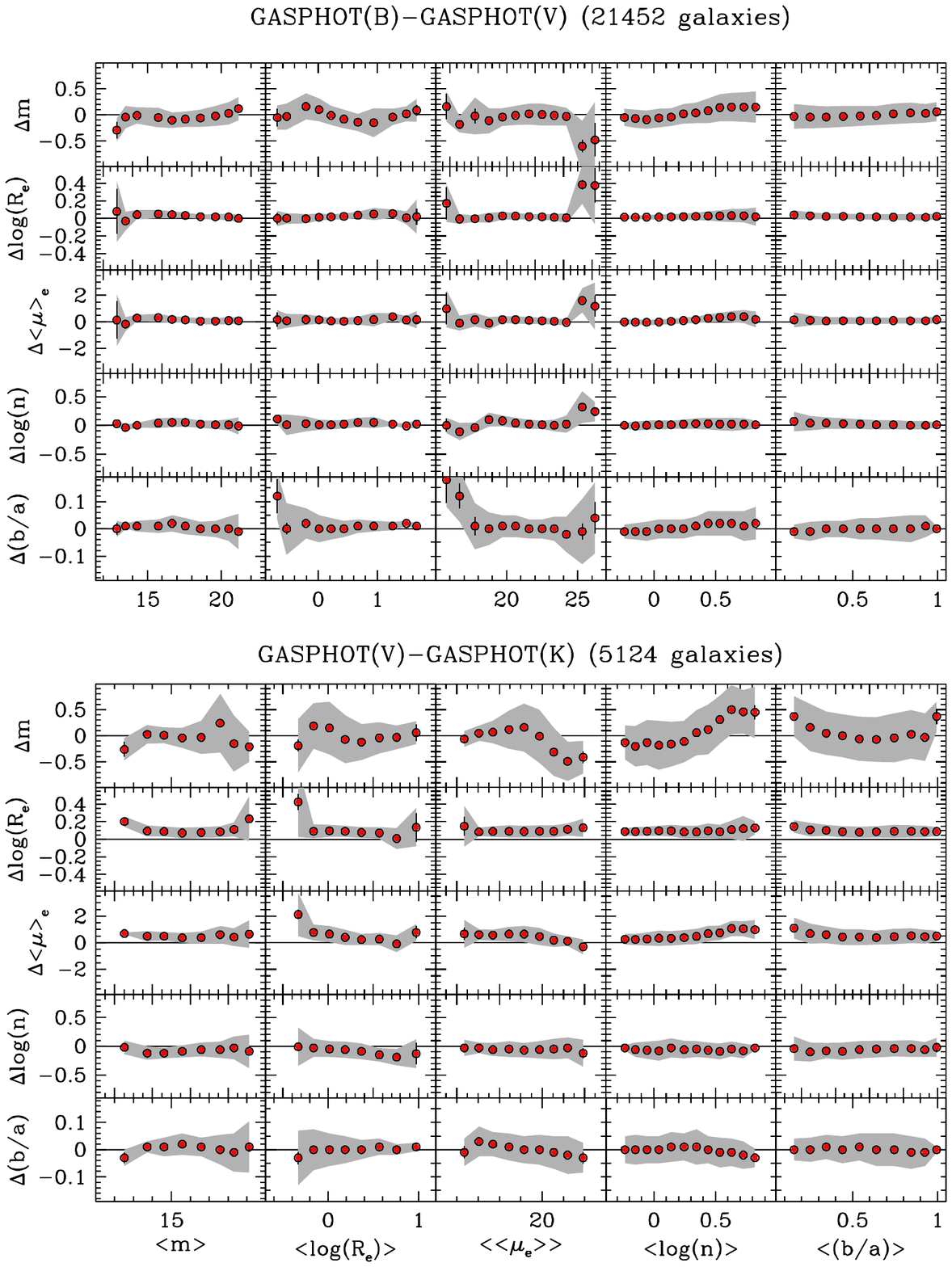}}
\vspace{-3truecm}
\caption{Comparison among various structural parameters obtained by
  GASPHOT in the different bands. The upper panels refer to the B-
  vs. V-band comparison, while the lower panels illustrate the V-
  vs. K-band comparison. The differences between parameters in
  different bands are plotted vs. the parameters themseves, averaged
  between the filters under comparison. The $\Delta$m and
  $\Delta$\muem differences are 'normalized' by removal of the average
  colors (see text). As in the previous figures, the error bars
    represent the uncertainties of the median values of the
    differences in each bin, while the shaded bands give the
    semi-inter quartile ranges of the distributions of the
    deviations.}
\label{Fig6}
\end{figure*}

Figures~\ref{Fig3} and \ref{Fig4} illustrate the differences between
surface photometry parameters for the GALFIT-GASPHOT and GIM2D-GASPHOT
comparisons, respectively. The surface photometry parameters used for
the comparisons are: the apparent total magnitude, the effective
radius in arcseconds, the effective average surface brightness, the
Sersic index and the axis ratio. Since we cannot a priori assume one
of the tools to give more reliable results than the others, in both
figures, the above parameters in abscissa are averaged between the
tools under comparison. The differences between the values found by
the tools are reported for each parameter on the ordinate, binned over
the whole set of parameters. Moreover, the error bars in the figures
represent the 1$\sigma$ uncertainties of the median values of the
differences in each bin, while the shaded bands give the
  semi-inter quartile ranges of the distributions of the deviations .

In Table~\ref{tab:Differences} the global median values and $r.m.s.$ of the
differences (GALFIT-GASPHOT) and (GIM2D-GASPHOT) are reported for the
same surface photometry parameters used in Figs.~\ref{Fig3}
and \ref{Fig4}.

From these figures and from Table~\ref{tab:Differences}
first we note that the general agreement is better and the scatter
smaller for the GIM2D-GASPHOT than for the GALFIT-GASPHOT
differences. This cannot be due to the different size of the two
samples (2581 vs.1684), since a larger scatter in the GALFIT-GASPHOT
plots is found even considering just early-type galaxies (plot not
shown), for which the sample size is greater for the GALFIT-GASPHOT
than for the GIM2D-GASPHOT comparison (1491 vs.1126). Thus, either we
should guess the intrinsic uncertainty to be larger for GALFIT than
for GIM2D or, alternatively, we could speculate about a sort of
environment driven, additional scatter, making the surface photometry
of galaxies less reliable in the cluster (GALFIT) than in the general
field (GIM2D) environment.

A second thing worthing to be noted in the plots is the dependence of
the scaling quantities differences ($\Delta m$, $\Delta log(R_e)$ and
$\Delta$\muem) on the galaxy scaling parameters themselves. In
particular, for large and bright galaxies, GASPHOT seems to produce
best-fit galaxy models brighter and larger than GALFIT and (just
marginally!) GIM2D. Since the same happens for the SExtractor-GASPHOT
comparison (see Fig.~\ref{Fig2}), we could be induced to
conclude this behaviour to be due to some size-drived bias of the
GASPHOT surface photometry (although an opposite tendency has been
noted before; see the first sentence of Sec.~\ref{Sec3}).  However, we
ruled out this conclusion on the basis of the following arguments:

(i) the above mentioned inability of \sext\ to extrapolate the
smoothly decreasing (high Sersic index) outer profiles of bright
galaxies is likely responsible of the size-driven magnitude
differences between \sext\ and GASPHOT, particularly for less deep
imaging, as in the case of the B- and (even more) K-band WINGS
imaging;

(ii) the agreement with the GIM2D photometry looks quite better than
for the other comparisons. In particular, the size-driven differences
are much less evident and some of the systematic differences present
in the GALFIT-GASPHOT comparison disappear, or even behave in the
opposite direction (see for instance $\Delta log(n)$ and $\Delta$\muem);

(iii) a natural attitude of GASPHOT to well represent the outer
luminosity profiles of large (halo-equipped) galaxies should be
expected, because of the GASPHOT tendency of overweighting the outer
regions of galaxies with respect to the other tools (see
Section~\ref{Sec3b4}). We think the most evident size-driven
differences between GALFIT and GASPHOT to be actually due to this
different weight allocation, which is particularly effective for
large, luminous galaxies. To this concern, it is interesting to note
that these systematic differences almost disappear if we exclude from
the sample the BCGs (Figure~\ref{Fig5}).

\subsubsection{Final remarks}\label{Sec3b6}

In general, the presence of systematic differences among the
parameters provided by different surface photometry tools should not
amaze anybody, since they are naturally expected because of the
different surface photometry techniques adopted by the tools. As
mentioned in Section~\ref{Sec2}, the GASPHOT algorithm is a sort of
hybrid between the 1D (equivalent luminosity profile fitting) and 2D
(full image fitting) approach.

Of course, pros and cons can be found for both approaches. As a
general rule, even if the 1D technique is unable to model either the
inner (seeing-affected) regions of flattened galaxies or possible
misalignments between different galaxy components, it has the
advantage of being less sensitive to the peculiar features of real
galaxies, since the elliptical isophotes are averaged over a large
number of pixels and their parameters (coordinates of the center,
semi-major axis, ellipticity and position angle) can be derived even
for very noisy and irregular shapes. On the contrary, the 2D approach
is fully equipped to handle the above mentioned issues (convolution of
seeing-affected regions of flattened galaxies and modeling of
misalignment between different galaxy components), but its sensitivity
to the galaxy peculiar features makes dangerous its blind (not
supervised) application to large galaxy samples, since it might
produce unrealistic results for a fraction of the sample. Roughly
speaking, the 1D approach is more robust, since it provides reasonable
results even in critical situations, while the 2D approach is suitable
for supervised, detailed modeling of well-sampled objects, even for
multi-component structures and in the very inner region of galaxies 
\citep{Haussler,PignFas06,Lotzetal2006,Blanton2003,Bershady2000}.

GASPHOT tries to exploit the robustness of the 1D fitting technique,
keeping at the same time the capability of dealing with PSF
convolution in the innermost regions typical of the 2D approach.
GASPHOT substantially reduces the amount of interaction for the user
and (mainly working in blind mode) is able to provide robust
estimates of the relevant global parameters for the hundreds of
galaxies typically found in wide/deep-field images.

\subsection{GASPHOT parameters in different bands}\label{Sec3c}

After having checked the results of GASPHOT against the alternative
tools GALFIT and GIM2D, in this section we compare among each other
the structural parameters obtained by GASPHOT in the B-, V- and
K-band.

In Figure~\ref{Fig6} the various structural parameters obtained by
GASPHOT in the different bands are compared as a function of the
parameters themselves, averaged between the filters under comparison.
In particular, the upper panels refer to the B- vs. V-band comparison,
while the lower panels illustrate the V- vs. K-band comparison. In the
plots comparing the total magnitude and the effective surface
brightness, in order to emphasize the trends of the relations, the
$\Delta$m and $\Delta$\muem\ differences are 'normalized' by
subtraction of the average colors, \ie the total magnitude
differences, averaged over the whole samples.

For all the photometric and structural parameters, the agreement
between the B- and V-band turns out to be fairly good. Moreover, no
significant trends are found, apart from a slight tendency of faint,
small galaxies and a more marked tendency of high Sersic index
galaxies to be redder. Instead, in the plots comparing the V- and
K-band GASPHOT parameters, the general agreement looks a bit worse
with respect to the previous case. In addition, various offsets and
trends are visible at a glance. In particular, compared with the
V-band, the structure of galaxies in the K-band shows (on average)
larger Sersic index, smaller effective radius and brighter effective
surface brightness, even after removal of the global galaxy color. In
addition, the tendency (already mentioned for the B- vs. V-band
comparison) of high Sersic index galaxies to be redder becomes much
more evident in the V- vs. K-band. All these facts are consistent with
the picture proposed by \citet[][see also
\citealt{Donofrio2013a}]{Donofrio2011}, in which, at increasing the
stellar mass (luminosity), early-type galaxies become on average
'older' (redder) and more centrally concentrated (higher Sersic
index). The dependence of the effective radius on the waveband has
been also discussed in \citet{Poggiantietal2013}.

\section{V-band structural parameters of galaxies in the WINGS clusters}\label{Sec4}

In this Section we briefly illustrate a few statistical properties of the
structural parameters of WINGS galaxies and some relations among them.
In order to produce the plots presented in this Section, the galaxy
sample described in Sec.~\ref{Sec3} has been further restricted to
include just the spectroscopic members of WINGS clusters for which we
have V-band GASPHOT surface photometry (3,131 galaxies). Relying on
the morphological classification obtained by MORPHOT, we divided this
sample in four, broad morphological types: (i) elliptical galaxies
(T=-5, 952 objects); (ii) S0 galaxies (-5$<$T$\le$0, 1478 objects);
(iii) early spirals (0$<$T$<$5, 593 objects); (iv) late spirals
(T$\ge$5, 108 objects).

\begin{figure}[h]
\resizebox{\hsize}{!}{\includegraphics{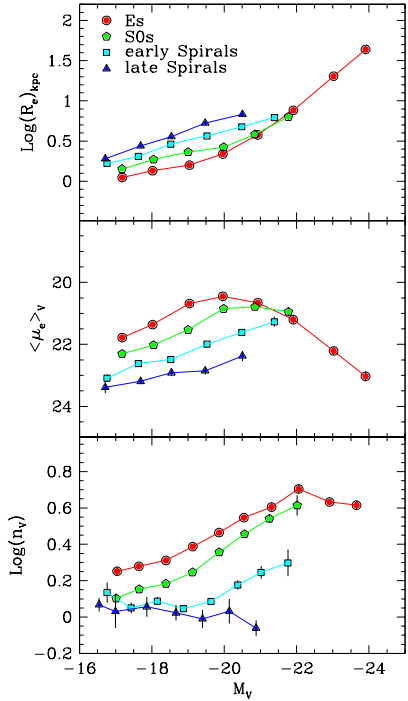}}
\vspace{-1truecm}
\caption{Effective radius (upper panel), mean surface brightness and
  Sersic index (lower panel) obtained by GASPHOT for the WINGS
  galaxies in the V-band as a function of the absolute magnitude for
  the four, broad morphological types. The mean values of the structural
  parameters are binned over the absolute magnitude. Ellipticals, S0s,
  early and late spirals are respectively represented by circles,
  pentagons, squares and triangles (red, green, cyan and blue in the
  electronic version).}
\label{Fig7}
\end{figure}

\begin{figure*}[h]
\resizebox{\hsize}{!}{\includegraphics[angle=-90]{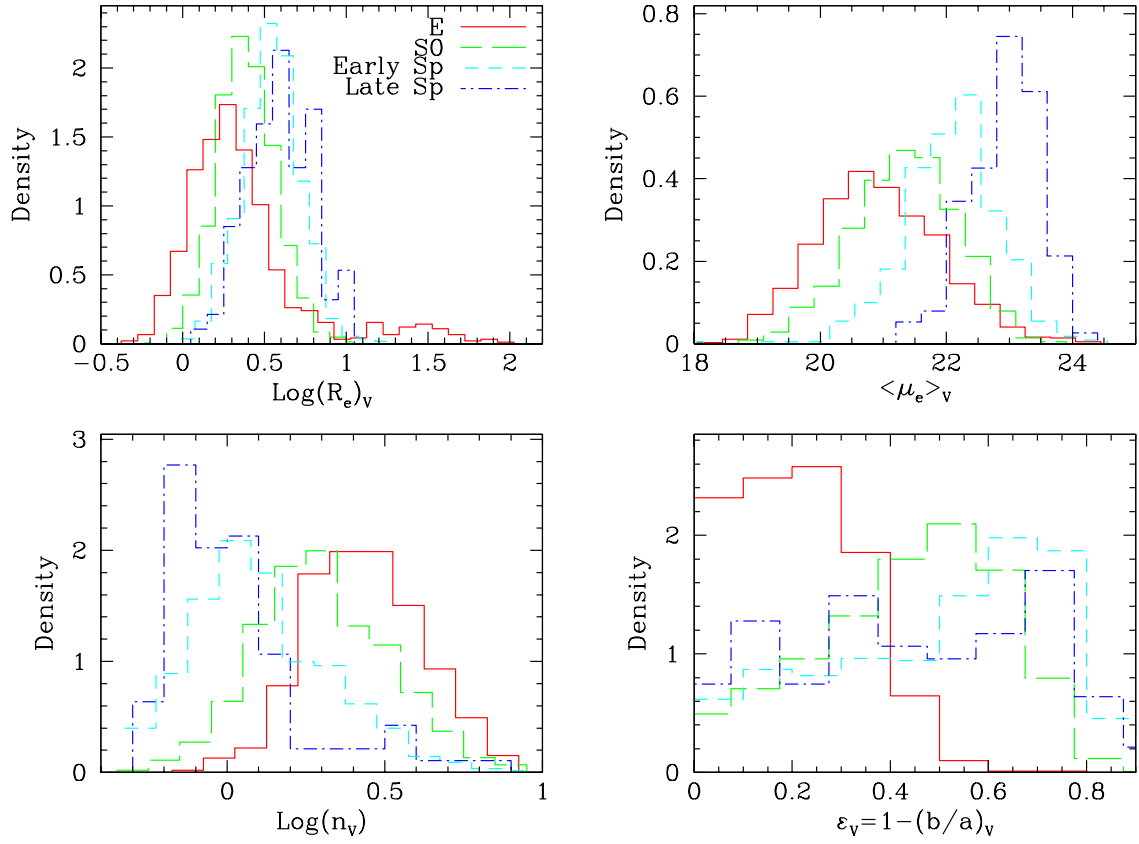}}
\caption{ Distribution of the V-band structural parameters derived by
  GASPHOT for the four, broad morphological types. The distributions
  are normalized to the area subtended by the histograms. Ellipticals,
  S0s, early and late spirals are respectively indicated by solid,
  long--dashed, short--dashed and dot-dashed lines (red, green, cyan
  and blue in the electronic version).}
\label{Fig8}
\end{figure*}

\begin{figure*}[h]
\resizebox{\hsize}{!}{\includegraphics{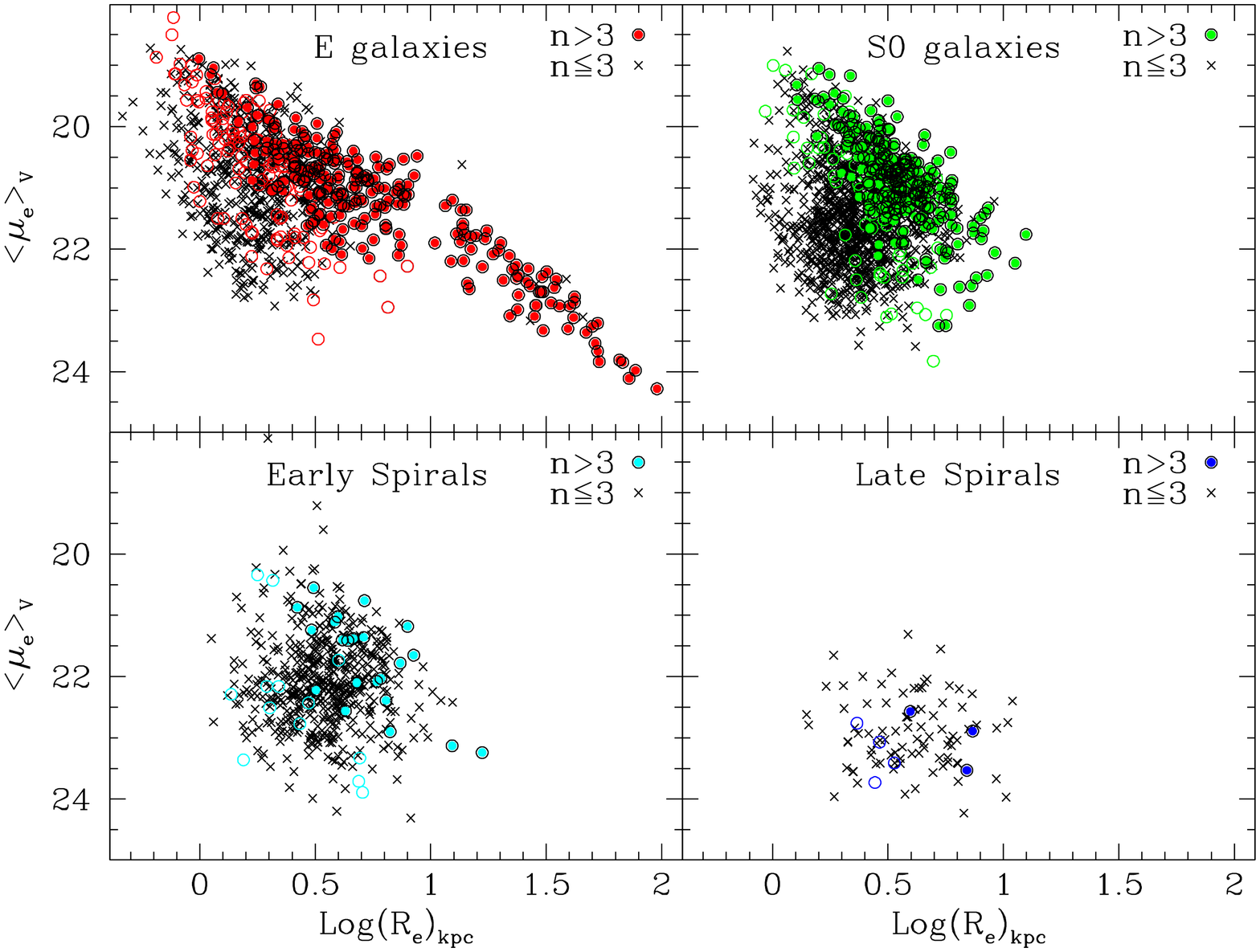}}
\caption{The plane \muem\ $- \log(R_e)$ for the galaxies of the four,
  broad morphological types and for two ranges of Sersic index (circles: n$>$3;
  crosses: n$\le$3). Open circles mark galaxies with n$>$3
  and isophotal area smaller than 10$^3$ pixels at 2.5$\sigma_{bkg}$}
\label{Fig9}
\end{figure*}

Figure~\ref{Fig7} illustrates how the structural parameters obtained
by GASPHOT for the WINGS galaxies in the V-band, depend on the
absolute magnitude for the four, broad morphological types. It shows
that, for a given absolute magnitude, the later the morphological
type, the lower the Sersic index and the larger the effective radius
and surface brightness. Moreover, at increasing the total luminosity,
the effective surface brightness decreases (with the notable exception
of the brightest Ellipticals), while both the Sersic index and the
effective radius increase, with the exception of the Sersic index of
late spirals, which slightly decreases at increasing the total
luminosity. Similar trends have been found in the recent literature
\citep{Trujillo2014,Bernardi2014}.

Note that the brightest Ellipticals show a strong overturning of the
surface brightness trend and a less pronounced (but still clear) break
of the Sersic index trend. The last feature, not present in the
classical relation discovered by \cite{Caonetal93} for the Virgo
cluster galaxies, has likely emerged here because of the much more
robust statistics on the BCGs. These features, together with the
marked increase of their size at increasing the total luminosity, are
consistent with the picture of BCGs as a separate class of objects,
distinct from normal Ellipticals and dominated by the cD galaxies
\citep{Capaccioli92,Fasano2010}.

In Figure~\ref{Fig8} the distributions of the same structural
parameters of Fig.~\ref{Fig7} (besides the ellipticity), normalized to
the area subtended by the histograms, are reported for the four, broad
morphological types. It is worth noting the remarkable continuity of
the distributions when moving from Elliptical galaxies towards later
types. To this concern, the ellipticity distributions constitute, in
some sense, an exception, since the flattening distribution of Es
looks quite different from those of any other morphological type, in
agreement with previous analyses \citep{FasanoVio,Fasanoetal93,Fasano2010}.

Finally, in Figure~\ref{Fig9} we plot onto the plane \muem\ $-
\log(R_e)$ the galaxies of the four, broad morphological types, in
turn divided in two subsamples, according to the Sersic index (full
dots: n$>$3; crosses: n$\le$3). The well known Kormendy relation
\citep[KR;][]{Kormendy77,Capaccioli92} seems to hold just for high Sersic
index Elliptical galaxies.  It looks much less evident (and with a
different slope) for low Sersic index Es and S0s, while it is not
present at all for spiral galaxies, at least when a single Sersic law
is used to represent their luminosity profiles.
The large scatter of the KR even for Es with n$>$3 turns out to be
reduced if we consider just galaxies with isophotal area larger than
10$^3$ pixels at 2.5$\sigma_{bkg}$ (full dots in Figure~\ref{Fig9}).

\section{Conclusions}\label{Sec5}

In this paper we have presented the B-, V- and K-band structural
parameters of the WINGS cluster galaxies with isophotal area larger
than 200 pixels at the threshold of 2.5$\sigma_{bkg}$ in each
band. The surface photometry has been obtained by means of the
automatic tool GASPHOT, which performs a simultaneous $\chi^2$
best-fitting of the major- and minor-axis growth curves of galaxies
using a single Sersic law convolved with a space-varying PSF. For each
cluster of the WINGS survey, GASPHOT produced catalogs for the
three photometric bands. The catalogs, available at the CDS,
provide, for each galaxy, the WINGS identification, the coordinates of
the galaxy center, the total magnitude, the effective radius, the
Sersic index, the axis ratios and a quality index flag related to the
goodness of each fit.

Thank to the database presented here, several thousands galaxies in
nearby clusters have now a robust characterization of their structural
properties.  These data have been already used in many papers of the
WINGS series \citep[see e.g.,][]{Donofrio2013a, Poggianti2013a,
  Poggianti2013IAUS, Donofrio2013b, Poggiantietal2013, Vulcani2012,
  Fasano2012, Vulcani2011a, Vulcani2011b, Bettoni2011, Donofrio2011,
  Ascaso2011, Valentinuzzi2010a, Fasano2010, Valentinuzzi2010b,
  Poggianti2009}.  The WINGS database, including the GASPHOT catalogs,
is useful for comparing the results of high
redshift surveys with the zero point reference frame of objects at low
redshifts. A complete description of this database can be found in
\cite{morettietal2014}.

The GASPHOT output has been tested through direct comparisons against
\sext\ (just for total magnitudes), GALFIT and GIM2D. The comparison
with GALFIT has been done in the \vf\ band using a subset of the WINGS
data, while that with GIM2D made use of the PM2GC data in the \bfilt\
band.  The agreement among GASPHOT and the above mentioned photometric
tools turned out to be generally good, apart from the tendency of
\sext\ to progressively underestimate the luminosity of large/bright
galaxies with respect to GASPHOT (see Sec.~\ref{Sec3a}). A similar
(less pronounced), size-driven bias seems to be present also when
comparing the total magnitudes from GASPHOT with those coming from
GALFIT. However, such a bias disappears if we remove the BCGs from the
comparison sample. This is likely due to the GASPHOT tendency of
overweighting the outer regions of galaxies with respect to the other
tools, this tendency being particularly effective for large, luminous
galaxies (see Sec.~\ref{Sec3b4}).

The uncertainties of the surface photometry parameters can be estimated
looking at the scatter of the differences among the values of the
parameters obtained using different tools. In particular, the average
uncertainty of the total luminosity is $\sim$12\% for both the
GALFIT-GASPHOT and GIM2D-GASPHOT differences.  Instead, the average
uncertainties of the other surface photometry parameters turn out to
be significantly lower for the GIM2D-GASPHOT than for the
GALFIT-GASPHOT differences. They range from 7\% to 20\% for the
effective radius and from 20\% to 30\% for both the effective surface
brightness and the Sersic index. These uncertainties are likely due
to various effects, as already discussed by \cite{PignFas06} and
\cite{Haussler}. Among them we stress the intrinsically different
``philosophy'' of the 1D approach followed by GASPHOT with respect to
the 2D approach followed by GALFIT and GIM2D.  We have seen that these
two methods have a different sensitivity to the peculiar features of
galaxies and behave differently in weighting the various (inner/outer)
galaxy regions.

The comparison among GASPHOT results in different wavebands shows a
fairly good agreement. Moreover, the trends observed in the colors
(especially the V-K) as a function of \muem\ and of the Sersic index,
are consistent with the picture proposed by \citet[][see also
\citealt{Donofrio2013a}]{Donofrio2011}, in which, at increasing the
stellar mass (luminosity), early-type galaxies become on average
'older' (redder) and more centrally concentrated (higher Sersic
index).

In conclusion, GASPHOT has proven to be effective in performing
automatic, blind surface photometry of galaxies in the intermediate
and low spatial resolution regime (ground-based, wide-field imaging at
low redshift or space-based imaging at intermediate redshift). In
these cases, GASPHOT is able to quickly provide robust structural
parameters of large galaxy samples.  We plane to use GASPHOT to obtain
the surface photometry of galaxies in the wide-field images of many
southern clusters taken with OmegaCam@VST in the framework of the
WINGS survey.

\begin{acknowledgements}
The thank Ignacio Trujillo for his help in the analysis of the WINGS galaxies with GALFIT.
\end{acknowledgements}

\bibliographystyle{aa}
\bibliography{biblio}

\end{document}